\documentclass[%
 reprint,
 showpacs,
 amsmath,amssymb,
 aps,
 prl,
 floatfix,
]{revtex4-1}
 \usepackage{amsmath}
 \usepackage{amsfonts}
 \usepackage{graphicx}
 \usepackage{epsfig}
 \usepackage{epstopdf}
 \usepackage{subfigure}
\usepackage{dcolumn}
\usepackage{epstopdf}
 \usepackage{hyperref}
 \hypersetup{
    unicode=false,          
    pdftoolbar=true,        
    pdfmenubar=true,        
    pdffitwindow=false,     
    pdfstartview={FitH},    
    pdftitle={title},    
    pdfauthor={D. G. Green},     
    pdfsubject={subject},   
    pdfcreator={D. G. Green},   
    pdfproducer={Producer}, 
    pdfkeywords={keywords}, 
    pdfnewwindow=true,      
    colorlinks=true,       
    linkcolor=blue,          
    citecolor=blue,        
    filecolor=blue,      
    urlcolor=blue           
}
\newcommand{\bi}{\begin{itemize}}
\newcommand{\ei}{\end{itemize}}
\newcommand{\bea}{\begin{eqnarray}}
\newcommand{\eea}{\end{eqnarray}}
\newcommand{\eqn}[1]{Eq.~(\ref{#1})}
\newcommand{\fig}[1]{Fig.~\ref{#1}}

\begin{document}
\title{Transverse Spreading of Electrons in High-Intensity Laser Fields} 
\author{D.~G.~Green}
\altaffiliation[Present Address: ]{Joint Quantum Centre (JQC) Newcastle/Durham, Department of Chemistry, Durham University, DH1 3LE, UK}
\email[\\]{dermot.green@balliol.oxon.org}
\author{C.~N.~Harvey}
\email[]{christopher.harvey@qub.ac.uk}
\affiliation{Centre for Plasma Physics, Queen's University Belfast, Belfast, BT7 1NN, UK}
\begin{abstract} 
We show that for collisions of electrons with a high-intensity laser, discrete photon emissions introduce a transverse beam spread that is distinct from that due to classical (or beam shape) effects.
Via numerical simulations, we show that this quantum induced transverse momentum gain of the electron is manifest in collisions with a realistic laser pulse of intensity within reach of current technology, and we propose it as a measurable signature of strong-field quantum electrodynamics.%
\end{abstract}
\maketitle
In recent years there has been steady increase in the powers and 
intensities of state of the art laser facilities (up to the current record of $2\times 10^{22}$\,Wcm$^{-2}$ \cite{Yanovsky:2008}).  
Numerous projects are now underway to ensure that this trend continues,
e.g., the Vulcan 10\,PW upgrade~\cite{Vulcan}, the ``Extreme Light Infrastructure'' (ELI) Facility~\cite{ELI} and the XCELS project~\cite{XCELS}, which aim to provide peak intensities of $10^{23}$--$10^{25}$\,Wcm$^{-2}$.
The development of such facilities has led to a renewed interest in probing strong field quantum electrodynamics (QED) using high-intensity laser fields~\cite{DiPiazza:2011tq, Heinzl:2008an}.
Examples of such QED processes include vacuum polarisation, pair production and cascades (see, e.g., Ref.~\cite{DiPiazza:2011tq} and references therein).

At intensities far below the onsets of these processes, however, the basic dynamics of an accelerated particle are strongly affected by the radiation it emits.  
These `radiation reaction' effects are thus of fundamental interest.
Moreover, understanding the dynamics is key to the accurate development of state-of-the-art QED-plasma `particle-in-cell' (QED-PIC) simulation codes that are expected to drive the experimental efforts~\cite{Bell:2008zzb,RidgersPRL,Sokolov:2010am, Elkina:2010up}.

In this Letter we study the dynamics of a beam of high energy electrons interacting with a high-intensity laser pulse.
Specifically, we consider electron-laser collisions in an intensity regime in which 
the quantum emission effects are important, but which is below the pair production threshold.
We show that through their transverse motion, the electron dynamics alone provide detectable signatures of strong field QED (see also \cite{Ilderton:2013dba_0,PhysRevX.2.041004}), and at intensities in reach of current technology.

Understanding the radiation back reaction is one of the most fundamental and oldest problems in electrodynamics (see, e.g.,~\cite{jackson,akhiezer,heitler,qed}).  
The common starting point for the classical approach is the Lorentz-Abraham-Dirac (LAD) equation, which results from the  solution of the coupled Lorentz and Maxwell's equations~\cite{Lorentz:1905,Abraham:1905,Dirac:1938nz}. 
This equation, however, suffers from notorious defects, e.g., unphysical runaway solutions.  
These can be somewhat circumvented by introducing certain approximations to reduce the LAD to a more applicable (albeit more approximate) form.  
The most well known of these is the perturbative approximation by Landau and Lifshitz (LL)~\cite{LLII}, valid when the radiative reaction force is much less than the Lorentz force in the instantaneous rest frame of the particle.  
It was shown to be consistent with QED to the order of the fine structure constant $\alpha$~\cite{Ilderton:2013dba_0, *Ilderton:2013dba_1, 0038-5670-34-3-A04}.  
We note that alternative classical equations of motion do exist (see, e.g., Refs.~\cite{O'Connell:2012ee,Sokolov:2009}, and for a summary, Ref.~\cite{Hammond:2010}).

Regardless, the validity of classical approaches in general decreases as the intensity increases and quantum effects become important.  
Indeed, there have been a number of studies investigating quantum effects on the dynamics of particles in strong laser fields~\cite{PhysRevE.81.036412,PhysRevX.2.041004,Neitz:2013,Bulanov:2013cga}. (For related theoretical studies of high energy electrons in crystal systems, see, e.g., \cite{Kirsebom2001274,Khokonov2004,PhysRevLett.89.094801}). 
The most recent of these showed, via the use of a kinetic formalism, a broadening in the (longitudinal) energy distribution of an electron bunch in a counter-propagating laser~\cite{Neitz:2013}, due to photon emission. 
In that work, however, the authors considered a regime in which transverse effects could be neglected.
Expressions for the QED tree-level amplitudes are well known~\cite{Ritus:1985}.
In a general collision, however, an electron will emit multiple times.
At high-intensity, multi-photon emission amplitudes are dominated by multiple incoherent single-photon emissions \cite{DiPiazza:2010mv}. 
Photon spectra for electrons in simple fields have recently been calculated numerically \cite{DiPiazza:2010mv}, but for arbitrary fields this becomes more difficult.
For the latter, numerical simulations based on strong-field QED provide an alternative approach, which we take here. 
Importantly, the QED simulations also provide a direct means of predicting the electron dynamics, which are the subject of this paper.

We begin with a discussion of the dynamics of a classical and quantum electron in the prototype plane-wave field.
Following this, we present results of numerical simulations of an electron colliding with a plane-wave laser, and in a more realistic setup, with a paraxial Gaussian laser.
The simulations implement methods similar to those used in the benchmark QED-PIC codes~\cite{Bell:2008zzb,RidgersPRL,Sokolov:2010am, Elkina:2010up}.
We show that the dynamics of electrons that emit discrete quanta are very different to those treated classically, via e.g., the LL equation, and that this leads to detectable signatures of strong field QED at intensities that will soon be available.
Finally, we conclude with a discussion on the relevance of the results. 

\paragraph*{Conventions:--}
As a test model we consider a plane wave field propagating in the $z$-direction described by the null wave vector $k^{\mu}=\omega(1,0,0,1)$, with central frequency $\omega$. (Throughout we adopt natural units where $\hbar=c=1$.)  Taking the field to be polarised in the perpendicular direction, we introduce the polarisation vector $\epsilon=(0,1,0,0)$, such that our basis vectors satisfy $k^2=k\cdot\epsilon=0$, $\epsilon^2=-1$. 
We define a dimensionless measure of field intensity in terms of the peak electric field, $a_0\equiv eE/m\omega$.  
The electromagnetic field tensor of the wave is taken to depend arbitrarily on the phase $\phi\equiv k\cdot x=\omega(t-z)$; $F^{\mu\nu}(\phi)=a_0f(\phi)f^{\mu\nu}$, where $f^{\mu\nu}=(k^\mu\epsilon^\nu-k^\nu\epsilon^\mu)$ and $f(\phi)$, satisfying $f(-\infty)=f(\infty)=0$, is a function describing the pulse.   
Finally, we introduce the quantity $\Omega\equiv k\cdot u=\omega(u^0-u^3)$ which is the laser frequency as `seen' by the electron.

\paragraph*{Classical case:--}
We begin by considering a classical electron moving in the test field.  
Let the initial momentum $p_0=mu_0$ when $\phi=\phi_0=-\infty$.  
The effects of radiation reaction (RR) can be accounted for via the Landau Lifshitz (LL) equation~\cite{LLII}, 
\bea
\frac{\Omega}{\Omega_0}\frac{dp^\mu}{d\phi}=
\Omega_0^{-1}F^{\mu\nu}p_\nu
+r_0\bigg(\mathcal{O}(F^2)\bigg), \label{eqn:LL}
\eea
where $r_0=(2/3)(e^2\Omega_0/4\pi m)$ is the coupling of the radiative correction terms. 
Taking $r_0\rightarrow 0$ gives the Lorentz force equation. 
Multiplying \eqn{eqn:LL} by $k$ gives~\cite{DiPiazza:2008} (see also~\cite{Harvey:2011dp}).
\bea
\Omega =\frac{\Omega_0}{1+r_0 a_0^2J},
\eea
where $J\equiv\int_{\phi_0}^\phi d\phi^\prime f^2(\phi^\prime )$.  
Thus we see that $\Omega$ decreases with time as the particle interacts with the laser.  
(Observe also that in the case without RR, $\Omega$ is conserved.)  
For a plane-wave field such as ours, the LL equation has recently been solved analytically~\cite{DiPiazza:2008}.
For the transverse momentum, the solution is found to be
\bea
p_{\rm LL,\perp}=\frac{\Omega}{\Omega_0}\left(p_{0,\perp}+m a_0 I
-ma_0 r_0 H\right),
\eea
where $I\equiv\int_{\phi_0}^\phi d\phi^\prime f(\phi^\prime )$ and $H(\phi)\equiv f(\phi) +a_0^2\int_{\phi_0}^\phi d\phi^\prime J(\phi^\prime) f(\phi^\prime)$.  Since the pulse function $f(\phi)$ is symmetric and finite, the integrals $I$ and $H$ taken over the entirety of the field are zero. 
The net transverse momentum then reduces to the product of the initial one, $p_{0,\perp}$, and the decaying prefactor $\Omega$. 
Thus in the plane wave case a classical particle {\it cannot} gain transverse momentum, it can only lose it due to RR. 
(Without RR the net change is zero.)
Note that in a more realistic field, to be discussed later, a classical particle can indeed gain transverse momentum. 
(In fact, for the special case of a classical particle in a bichromatic field, such a gain has been proposed as a method of controlling the electron dynamics \cite{2013arXiv1306.3328T}.)
In the regime we consider, however, the quantum effects dominate and are clearly distinguishable.

\paragraph*{Including Compton scattering:--}

To consider quantum effects it is instructive to introduce the dimensionless and invariant  `quantum efficiency' parameter $\chi_e\equiv \sqrt{(F^{\mu}_{\phantom{\mu}\nu}p^\nu)^2}/m^2\sim\gamma E/E_{\rm cr}$, where $E_{\rm cr}=1.3\times10^{16}$\,Vcm$^{-1}$ is the QED `critical' field (`Sauter-Schwinger' field)~\cite{QEDcriticalfield1,*QEDcriticalfield2,*QEDcriticalfield3}.  
It can be interpreted as the work done on the electron by the laser field over the distance of a Compton wavelength.  
When $\chi_e\gtrsim1$ quantum effects dominate and pair production can occur.  
Thus in order to study quantum emission processes (viz.~Compton scattering) cleanly, as we do here, one should be in a regime where $a_0$, $\gamma\gg 1$, such that quantum effects play a role, but have $\chi_e\lesssim 1$, so that pair production can be neglected.  
In the limit $a_0\gg 1$ the size of the radiation formation region is of the order $\lambda/a_0\ll\lambda$, where $\lambda=2\pi/\omega$ is the laser wavelength~\cite{Ritus:1985}.   
Thus the laser varies on a scale much larger than the formation region and so can be approximated as locally constant and crossed, allowing us to determine the probability of photon emission using the differential rate~\cite{Ritus:1985}
\begin{eqnarray}\label{eqn:dGam}
{d\Gamma}&=&\frac{\alpha m}{\sqrt{3}\pi\gamma\chi_e}
\left[\left( 1-\eta+\frac{1}{1-\eta}\right) K_{2/3}(\tilde{\chi})\right.\nonumber\\
&&\left.-\int_{\tilde{\chi}}^\infty dx K_{1/3}(x)\right] {d\chi_\gamma},
\end{eqnarray}
where $K_\nu$ is the modified Bessel function of order $\nu$, 
$\eta\equiv \chi_{\gamma}/\chi_e$, $\tilde{\chi}\equiv 2\eta/\left[3\chi_e\left(1-\eta\right)\right]$, 
and we have introduced the analogous invariant parameter $\chi_\gamma\equiv \sqrt{(F^{\mu}_{\phantom{\mu}\nu}\kappa^\nu)^2}/m^2$ for the emitted photon with momentum $\kappa^\nu$.  
Note that although $d\Gamma$ diverges at small $\chi_{\gamma}$, the total differential probability of photon emission (i.e., of any $\chi_{\gamma}$), $dW = \Gamma dt$, where $\Gamma \equiv\int_0^{\chi_e}d\Gamma$, is finite (see also~\cite{Ilderton:2012qe}). 

In general the electron will radiate multiple times as it interacts with the laser field.
At high-intensity, multi-photon emission amplitudes are dominated by
multiple, incoherent single-photon emissions \cite{DiPiazza:2010mv}, each described by \eqn{eqn:dGam}.
The description of the interaction in terms of the invariants $\chi_e$ and $\chi_{\gamma}$ is then particularly instructive because the conservation law $\chi_e'=\chi_e-\chi_{\gamma}$ holds, where $\chi_e$, $\chi_e'$ are the initial and final electron invariants, respectively~\cite{Ritus:1985}. 

We have developed the single-particle code `{\tt SIMLA}' \footnote{Further details of {\tt SIMLA} can be requested from the authors.} that calculates the trajectory of an electron undergoing Compton scattering in an arbitrary background field.
Briefly, the code implements a classical particle pusher that propagates electrons through the field via the Lorentz (or LL) equation over discrete spatial and temporal grids. 
The emission process is implemented via statistical routines similar to those in a number of recently developed particle-in-cell codes for the modeling of QED cascades (see, e.g.,~\cite{Bell:2008zzb,Sokolov:2010am, Elkina:2010up}).  
(Similar theory and methods are employed in studies of high energy particle beams interacting with crystals~\cite{Kirsebom2001274,Khokonov2004,PhysRevLett.89.094801}.)  
At each time step a uniform random number $r\in[0,1]$ is generated, and emission deemed to occur if the condition $ r\leq \Gamma dt$ is satisfied, under the requirement $\Gamma dt\ll1$. 
Similar event generators have been used in Refs.~\cite{Elkina:2010up} and \cite{Duclous}.
Note that during the simulation $d\Gamma$ (and thus $\Gamma$) is a time-dependent quantity owing to the effect of the temporally varying laser pulse and electron motion. 
Given an emission event, the photon $\chi_{\gamma}$ is determined as the root of the sampling equation $\zeta={\Gamma(t)}^{-1} \int_{0}^{\chi_{\gamma}}d\Gamma(t)$, where $\zeta$ is a uniform random number $\zeta\in[0,1]$~\footnote{In practice, the integral is performed from a lower limit $\varepsilon \sim 10^{-5}$, rather than zero. 
The emission of soft photons of energy below this cut off does not appreciably affect the electron dynamics (see, e.g., Ref.~\cite{Duclous}).}.
Next, we calculate the photon momentum from $\chi_{\gamma}$ assuming that the emission is in the direction of motion of the electron.
This is valid for $\gamma\gg 1$, since in reality the emissions will be in a cone of width $\gamma^{-1}$~\cite{jackson, Harvey:2009ry}. 
Finally, the electron momentum is updated and the simulation continues by propagating the particle via the Lorentz equation to the next time step.

\paragraph*{Dynamics of a quantum electron:--}
Despite the discrete nature of emission, the dynamics in the quantum case can be described in a similar manner to the classical one.
Specifically, we assume once again that the particle enters the field with momentum $p_0$ at $\phi=\phi_0$.  It then propagates forward according to the Lorentz force until it emits a photon of momentum $\kappa_1$ when $\phi=\phi_1$, leaving it with momentum $p(\phi)=p(\phi_1)-\kappa_1$.  This process continues through the subsequent emissions of $n$ photons giving the equation of motion
\bea
p_{\rm QED}&=&p_0+m a_0I\epsilon-\sum_{i=1}^{n}\theta(\phi_i)\kappa_i\nonumber\\
&+&a_0k\sum_{i=0}^{n}\Omega_i^{-1} I_i
\left(
 \epsilon\cdot p_i
+\frac{1}{2}m a_0 I_i
\right)
,\label{eqn:pQED}
\eea
where $\theta(\phi)$ is the Heaviside function and $I_i (\phi)\equiv\int_{\phi_i}^{\phi_{i+1}} d\phi^\prime \theta(\phi) f(\phi^\prime)$.
If we consider just the transverse motion we find
\bea
p_{\rm QED,\perp}=p_{0,\perp}+ma_0 I -\sum_{i=1}^{n}\theta(\phi_i)\kappa_{i,\perp},
\eea
where again $I=0$ if taken over the entirety of the field.
Thus the (quantum) corrections to the transverse momentum come solely from the recoil of the emitted photons.  
The stochastic nature of these emissions means that the electron will perform a random walk in transverse momentum space.  
The result is that a particle can gain transverse momentum, unlike in the classical case where, for a plane wave field, it can only lose it.
Another point of note is that the frequency $\Omega$ also now changes discretely rather than continuously.  
Therefore a particle will initially see a laser of frequency $\Omega_0=k\cdot u_0$; this quantity will be conserved (as is consistent with the Lorentz force dynamics) until the first emission, after which it will see a frequency of $\Omega_1=k\cdot(p_0-\kappa_1)/m$, etc.  
However, since an experiment would likely consist of many electrons radiating at different times, one would expect this signal to be `washed out' in any observed spectra. 
In terms of understanding the particle dynamics, the effect of the discretely changing frequency in Eq.~(\ref{eqn:pQED}) is that the longitudinal momentum is dependent not just on the photon momentum, but also on the value of the phase when the photon is emitted.  
Thus in contrast to the transverse motion, the impact of discrete photon emissions on the longitudinal motion is somewhat obscured, although a recent work~\cite{Neitz:2013} has shown that there will be a longitudinal spreading of an electron bunch in a laser field (see also~\cite{Bulanov:2013cga, PhysRevX.2.041004}).
\begin{figure}[t!]
\includegraphics*[width=1\columnwidth]{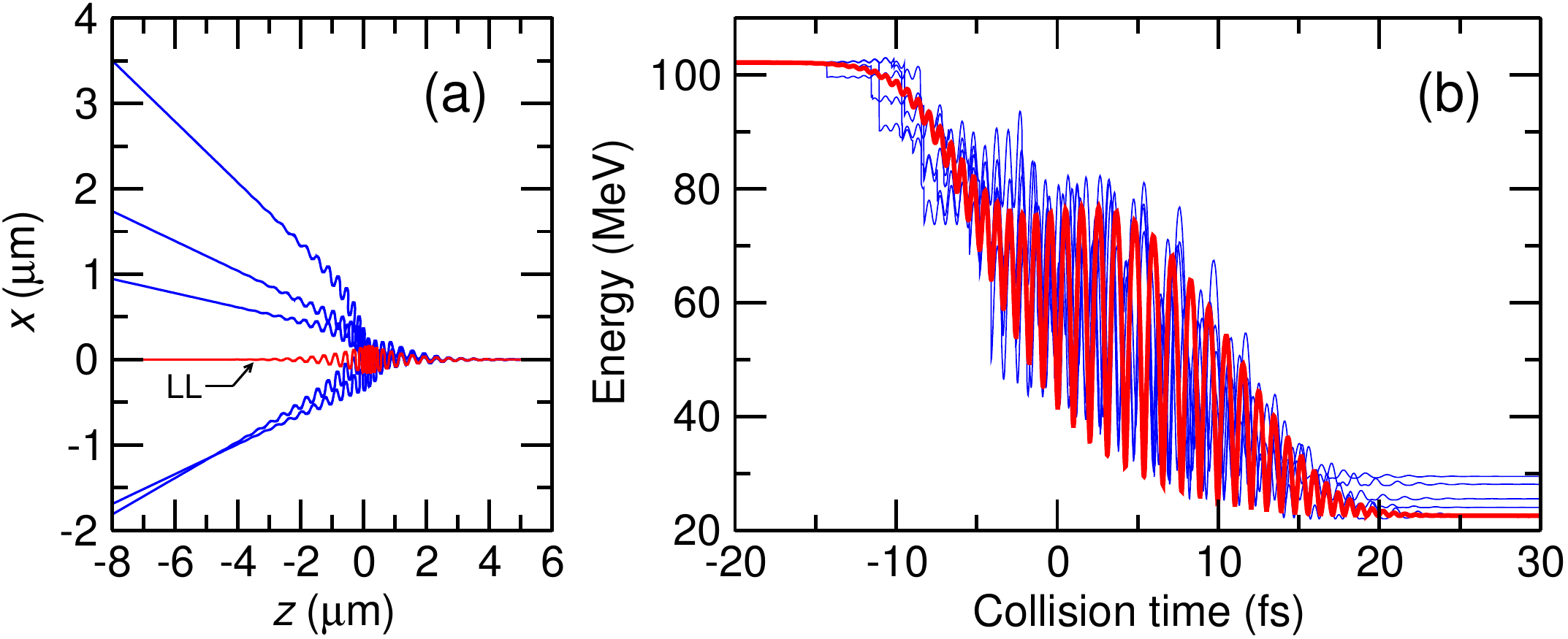}
\caption{Simulation results for electrons with initial energy of $102.2$\,MeV ($\gamma=200$) colliding head-on with a 30\,fs FWHM Gaussian plane-wave field of intensity $a_0=150$ and wavelength $\lambda=0.8\,\mu$m propagating in the positive $z$ direction.
LL [\eqn{eqn:LL}], red lines; sample of five QED (simulated) electrons, blue lines:
(a) trajectories;
(b) energies of electrons in (a). 
The electrons are timed such that they would enter the peak of the pulse at $t=z=0$, were they not affected by the field.
\label{fig:QEDpw} }
\end{figure}

In \fig{fig:QEDpw} we compare the QED simulation results with the solution of the classical LL equation, for head-on collisions of electrons of initial energy $102.2$\,MeV ($\gamma=200$) with a linearly polarised plane wave with a Gaussian time envelope of 30\,fs FWHM, and of peak intensity $a_0=150$ ($9.5\times 10^{22}$\,Wcm$^{-2}$).  
For these parameters $\chi_e\lesssim0.1$. 

As discussed, the LL electrons radiate continuously and, although they oscillate in the field, they return to the field axis $x=0$ after the collision.
Figure \ref{fig:QEDpw}\,(a) shows that, in contrast, the QED electrons can acquire significant final transverse motion.
In fact, we find that for 2500 simulated collisions the distribution in the final `opening' angle $\theta\equiv\arctan(x/z)$ is well described by a Gaussian of FWHM = 29.8 degrees.
Figure \ref{fig:QEDpw}\,(b) shows the energy of the electrons with collision time.
The discrete reductions in the energy of the quantum electrons are evident. 
The LL result is mainly contained within those of the QED electrons, falling to the lower end of the energy range at the end of the interaction.
(This is to be expected since the classical formula allows for the emission of photons with energy greater than the electron energy~\cite{DiPiazza:2010mv}.) 

\paragraph*{Realistic setup:--}
\begin{figure}
\includegraphics*[width=1\columnwidth]{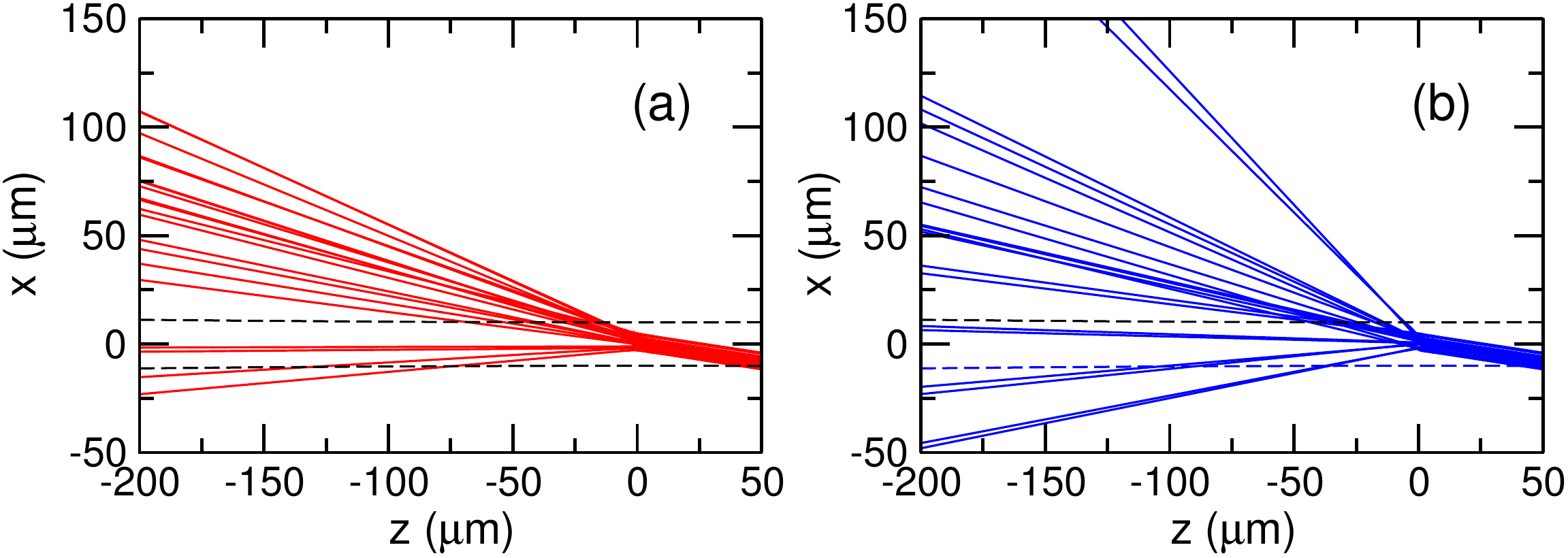}%
\caption{Sample of 20 simulated electron trajectories in a paraxial Gaussian laser beam of intensity $a_0=150$ ($9.5\times 10^{22}$\,Wcm$^{-2}$), wavelength of 800\,nm, waist radius of 10\,$\mu$m and duration 35\,fs (dashed line shows laser profile).  
Initial electron energies distributed around 255\,MeV ($\gamma_0=500$) according to a Gaussian of FWHM 0.52 MeV.  
Electrons incident at 10 degree angle.  (a) classical (LL); (b) QED electrons.\label{fig:QEDGauss} }
\end{figure}
To investigate whether these features are detectable experimentally, we now consider a more realistic setup. 
A typical source of electrons in such experiments is from a linear accelerator, e.g., the ELBE accelerator at the Forschungszentrum Dresden-Rossendorf in Germany~\cite{Arnold:2007}.  
We assume that the high charge mode beam is accelerated to $\gamma=500$ and the normalised transverse emittance of 2.5mm mrad is preserved.  The beam is then focused to 2.5\,$\mu$m diameter FWHM at the interaction point. Even taking into account the space-charge effects the beam divergence is still much smaller than the angular spread after laser-beam interaction, so for simplicity we assume a parallel incoming beam in the simulation.
We further assume the beam to be of high quality with a small energy spread ($\Delta\gamma/\gamma_0 =10^{-3}$ is feasible with the aforementioned facility~\cite{Heinzl:2009nd}.)
For our laser field we use a paraxial Gaussian beam defined to fifth order in the expansion parameter~\cite{Salamin:2002dd}, with a wavelength of $800$\,nm and beam waist $w=10\,\mu$m.  
This corresponds roughly to what is expected to be achievable at the future facilities referred to in the introduction.  
Finally, the electron beam is assumed to collide with the laser pulse at an angle of 10 degrees.
For these parameters $\chi_e\lesssim0.3$.

\begin{figure}[t!]
~\includegraphics*[width=1\columnwidth]{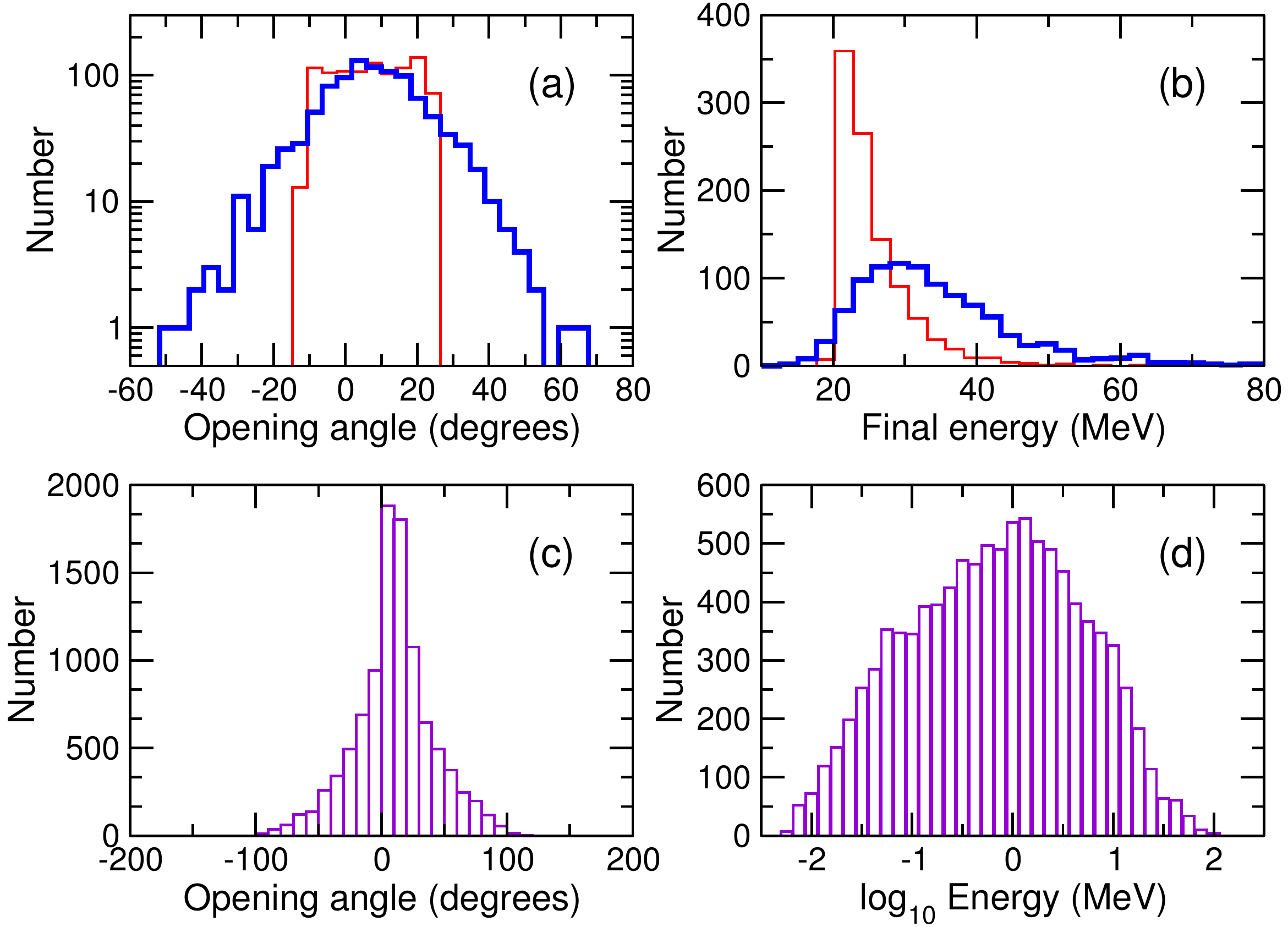}\\
\caption{
Simulation results for a bunch of 1000 electrons initially spatially distributed according to a Gaussian of $2.5\,\mu$m FWHM and satisfying an initial energy distribution centred around 255\,MeV ($\gamma_0=500$) according to a Gaussian of FWHM 0.52 MeV. 
Laser parameters as \fig{fig:QEDGauss}.  Upper panels: (a) final electron opening angles $\theta\equiv\arctan(x/z)$, (b) final electron energies. (Thin, red lines: classical (LL); thick, blue lines: QED). Lower panels: (c) opening angles and (d) energies of the emitted photons for the QED simulation.  \label{fig:Gaussian_angle} }
\end{figure}

In \fig{fig:QEDGauss} we show the trajectories of a random sample of 20 electrons for both the classical and QED cases.  
It is known that for a realistic field such as ours, the combination of finite size effects and classical RR will also cause a certain amount of spreading of the electron beam~\cite{Harvey:2011mp}.  However, it is clear from the plots that these effects are not enough to mask the QED-induced transverse spread predicted from the plane wave analysis.  
Indeed, the quantum simulation is markedly different from its classical counterpart.  
To investigate further, we consider the statistics of a bunch of 1000 electrons following the same initial distribution as before.  
The final opening angles and energies are shown in \fig{fig:Gaussian_angle}.  
It can be seen that the stochastic nature of the quantum photon emissions causes the angular spread of the electrons to increase by roughly a factor of two. 
This distinctive broad QED `shoulder' should be experimentally detectable.
Finally, \fig{fig:Gaussian_angle}\,(b) shows that again, the QED electrons experience a broadening in their energy distributions, consistent with~\cite{Neitz:2013,PhysRevX.2.041004}.
(Note that in both cases, the particles that glance off the fringes of the pulse can experience a net gain in energy, see, e.g.,~\cite{Salamin:2002dd}.)
Finally, in \fig{fig:Gaussian_angle}\,(c) and (d) we show the properties of the emitted photons for the QED simulation.  
It can be seen that, as expected, the majority of the photons are emitted roughly in the direction of the electron beam axis.  The photon energies follow a synchrotron-like spectral distribution \cite{jackson}, with a peak around 0.02 MeV.
For highly relativistic particles in intense fields the calculation of the classical spectrum is computationally expensive \cite{Reville}, and is beyond the scope of the paper.
However, it is possible to make a qualitative comparison with the classical spectra.
We find that the energy of the classical electron drops to $\sim 50$\,MeV by the time the particles reach the most intense part of the field, where most emissions will occur.  In the classical case the intensity of the emitted radiation rapidly decreases \cite{Sprangle} after the critical harmonic $n\sim 3a_0^3/4$, with frequency $\omega_n=4\gamma^2 n\omega/(1+a_0^2/2)$ (see e.g.~\cite{Harvey:2009ry}).  In our case $\omega_{\textrm{crit}}=13.9$\,MeV, which is consistent with the fall-off in the QED spectrum shown in \fig{fig:Gaussian_angle}~(d).
\paragraph*{Conclusions and outlook:--}
We have studied the dynamics of an electron in a high-intensity laser field.
Using a plane wave test model we first showed formally that when a particle emits discretely it takes a random walk in transverse momentum space, and can thus gain, as well as lose, transverse momentum.
This is in contrast to the classical case where, in a plane wave field, a radiating particle can only experience a net loss of transverse momentum.
Finally, we performed a numerical study of a beam of electrons interacting with a paraxial beam model of a laser and showed that the effect is also clearly exhibited in this more realistic setup.  
The results show that the electron dynamics alone should provide a measurable signature of QED at intensities only marginally higher than currently available.

\acknowledgements
We thank Michael Geissler, Thomas Grismayer, Tom Heinzl, Anton Ilderton, Constantin Klier, and Matt Zepf for useful discussions.
The authors were funded by the EPSRC, UK, and both contributed equally to this work.


%

\end{document}